\documentclass[sigconf,natbib=true]{acmart}

\AtBeginDocument{%
  }

\copyrightyear{2025}
\acmYear{2025}
\setcopyright{cc}
\setcctype{by}
\acmConference[CIKM '25]{Proceedings of the 34th ACM International Conference on Information and Knowledge Management}{November 10--14, 2025}{Seoul, Republic of Korea}
\acmBooktitle{Proceedings of the 34th ACM International Conference on Information and Knowledge Management (CIKM '25), November 10--14, 2025, Seoul, Republic of Korea}
\acmDOI{10.1145/3746252.3760822}
\acmISBN{979-8-4007-2040-6/2025/11}

\usepackage{multirow}

\begin{document}

\title{+VeriRel: Verification Feedback to Enhance Document Retrieval for Scientific Fact Checking}

\author{Xingyu Deng}
\email{xdeng37@sheffield.ac.uk}
\affiliation{%
  \institution{University of Sheffield}
  \city{Sheffield}
  \country{UK}}

\author{Xi Wang}
\email{xi.wang@sheffield.ac.uk}
\affiliation{%
  \institution{University of Sheffield}
  \city{Sheffield}
  \country{UK}}

\author{Mark Stevenson}
\email{mark.stevenson@sheffield.ac.uk}
\affiliation{%
  \institution{University of Sheffield}
  \city{Sheffield}
  \country{UK}}

\begin{abstract}
Identification of appropriate supporting evidence is critical to the success of scientific fact checking. However, existing approaches rely on off-the-shelf Information Retrieval algorithms that rank documents based on relevance rather than the evidence they provide to support or refute the claim being checked. 
This paper proposes \textit{+VeriRel} which includes verification success in the document ranking. Experimental results on three scientific fact checking datasets (SciFact, SciFact-Open and Check-Covid) demonstrate consistently leading performance by \textit{+VeriRel} for document evidence retrieval and a positive impact on downstream verification. 
This study highlights the potential of integrating verification feedback to document relevance assessment for effective scientific fact checking systems. It shows promising future work to evaluate fine-grained relevance when examining complex documents for advanced scientific fact checking.

\end{abstract}

\begin{CCSXML}
<ccs2012>
   <concept>
       <concept_id>10002951.10003317.10003371</concept_id>
       <concept_desc>Information systems~Specialized information retrieval</concept_desc>
       <concept_significance>500</concept_significance>
       </concept>
 </ccs2012>
\end{CCSXML}

\ccsdesc[500]{Information systems~Specialized information retrieval}

\keywords{Evidence retrieval, Scientific fact checking, Document ranking}

\maketitle

\section{Introduction}\label{sec:introduction}

Automated scientific fact checking is the process of verifying or refuting scientific claims using peer-reviewed research as evidence. The problem has become increasingly important given the rapid growth rate of scientific knowledge and the associated challenge of keeping up to date. Increasing interest in the problem is demonstrated by the development of novel approaches and release of datasets, e.g., \citep{kotonya-toni-2020-explainable-automated,wadden-etal-2020-fact,sarrouti-etal-2021-evidence-based,mohr-etal-2022-covert,wadden-etal-2022-scifact,wang-etal-2023-check-covid}. 

A critical stage of the fact checking process is the identification of evidence against which a claim can be evaluated \cite{10.1145/3731120.3744614}. This problem normally involves searching for documents containing this evidence within a large corpus, such as a repository of scientific publications. A distinction can be made between ``evidential'' and semantically relevant documents. The first contains evidence regarding the claim's correctness, while the second includes information related to the claims, but may or may not contain useful evidence. For example, given the claim ``\textit{Ibuprofen is frequently used to treat headaches in COVID-19 patients.}'', the statement ``\textit{Ibuprofen can be used for headache treatment}'' is relevant but not evidential. However, ``\textit{Paracetamol is the most commonly used medicine for COVID-19 for any symptom}'' is both relevant and evidential. Previous work has demonstrated that retrieving relevant but non-evidential documents can be harmful to the fact checking process. For example, \citet{sauchuk2022role} reported that combining evidential documents identified by a perfect retriever with a single relevant but non-evidential document produced a 17.2\% drop in fact checking performance.

Previous approaches to identifying evidential information have focused on sentence-level evidence selection. \citet{zheng-etal-2024-evidence} employed joint optimisation but their tight coupling of retrieval and verification risks overfitting and limited generalisability. Others explored sequential retriever refinement \cite{zhang-etal-2023-relevance,hu2023read} which requires gold sentence-level evidence for supervision, limiting applicability beyond annotated data.
A significant limitation of previous approaches is that they assume the documents containing the evidential information are already provided, ignoring the fact that this information is not normally available. Despite its importance, integrating verification feedback into document-level retrieval remains an unexplored problem.

This study addresses this gap to advance document retrieval for scientific fact checking using joint estimation of semantic relevance and downstream verification success of documents. The proposed approach is trainable on arbitrary claim–document pairs using automatically derived verification feedback, thereby enabling broad applicability and scalable training.
The main contributions of this paper are:

\noindent\textbf{\small$\bullet$} Propose \textit{+VeriRel}, a trainable reranker that effectively integrates verification feedback for document ranking. Results show consistent leading performance compared to state-of-the-art baselines across three datasets.

\noindent\textbf{\small$\bullet$} Explore the factors that can affect the scalability and domain generalisability of \textit{+VeriRel}, which suggests a novel way to train robust document reranking models for scientific fact checking.

\begin{figure*}[ht]
\centering
  \begin{minipage}{0.9\textwidth}
    \includegraphics[width=\linewidth]{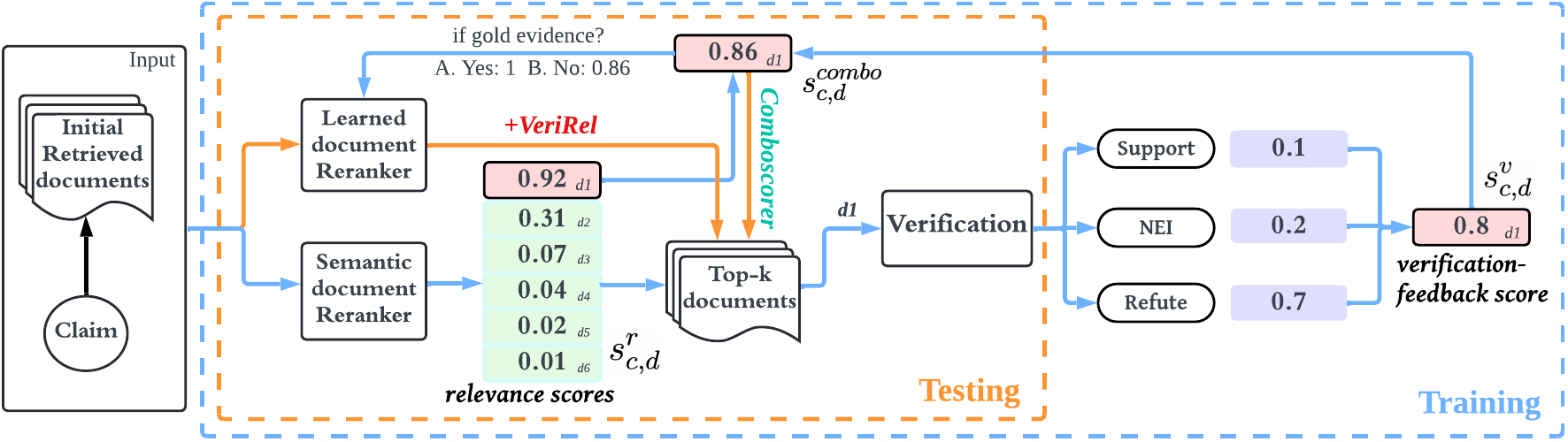}
    \caption{Pipeline to leverage verification feedback to enhance document retrieval. 
    The blue pathway presents the flow of producing joint scores for \textit{+VeriRel} model training.
    The orange pathway shows the test of ranked documents (1) by calculated scores and (2) by trained \textit{+VeriRel}.}
    \label{fig:VeriFED}
  \end{minipage}
  \Description{The figure about pipeline of proposed +VeriRel, including training and testing in this study}
\end{figure*}
\section{Methodology}
\subsection{Task definition}\label{ssec:problem_statement}

Scientific fact checking systems aim to assess a claim, $c$, based on the information contained within a corpus of documents, $D$. Given this information, a fact checking system, $FC$, identifies a set of documents, $D' \subset D$, and predicts a label for each by a classification model, with labels drawn from a pre-defined set, $\mathcal{L}$, i.e. 

\begin{equation}
\label{pipeline_euqation}
  FC(c,D) \rightarrow \left\{ (d, l): d \in D', l \in \mathcal{L} \right\}
\end{equation}

Each label is an assessment of how the information within the document relates to the veracity of the claim. A commonly used a set labels is $\mathcal{L} = \{ {\tt SUPPORT}, {\tt REFUTE}, {\tt NOT\;ENOUGH\;INFORMATION} \}$.

This process is normally carried out in a two-stage pipeline: evidence retrieval and verification. The first of these, $Retrieval(\cdot)$, treats $c$ as a query 
and retrieves $D'$, the top $k$ ranked documents from $D$:
\begin{equation}
\label{equa:retrieval_equation}
  D' = Retrieval(c,D,k)
\end{equation}

The verification stage is then applied to each document $d \in D'$ to determine its label: 
\begin{equation}
\label{equa:verifier_equation}
V(c,d) = \underset{l \in \mathcal{L}}{argmax}(P(l | c, d))
\end{equation}

where $P(l | c, d)$ is an estimate of the probability of label $l$ given the claim and document. The verifier chooses the label with the highest probability score.

\subsection{+VeriRel}
\label{ssec:+verirel}
Document retrieval typically involves two steps: initial retrieval for high recall, followed by reranking to identify top-relevant documents~\citep{hambarde2023information}. This pipeline allows scientific fact checking systems to obtain semantically relevant documents effectively. However, as noted in Section~\ref{sec:introduction}, semantic relevance alone may introduce non-evidential noise. Since the verification stage 
estimates how likely a document supports, refutes, or lacks information for a claim, it is therefore hypothesised that \textit{incorporating verification feedback into retrieval can enhance the identification of truly evidential documents}.

\textbf{\textit{+VeriRel}} combines semantic relevance and verification feedback to improve document ranking. A document’s verification usefulness as feedback $s^{v}_{c,d}$ is defined as the sum of support and refute probabilities produced by the verifier (Eq.~\ref{equa:verifier_equation}), while the predicted probability of the {\tt NOT\;ENOUGH\;INFORMATION} label is intentionally ignored: 
\begin{equation}
\label{equa:SA_euqation}
s^{v}_{c,d} = P({\tt SUPPORT}|c, d) + P({\tt REFUTE}|c, d)
\end{equation}

\noindent Semantic relevance, $s^r_{c,d}$, is computed using an existing ranking model. The final ranking score is a weighted combination that approximates a non-learned ranking strategy to take account of both the semantic and evidential value of a document:
\begin{equation}
\label{ARF_equation}
s^{combo}_{c,d} = \alpha \times s^{v}_{c,d} + (1 - \alpha) \times s^{r}_{c,d},\ \alpha \in [0, 1]
\end{equation}

However, this formulation requires explicit verifier outputs for every document which limits its practicality for large-scale deployment. To address this, \textit{+VeriRel} approximates the joint signal:
\begin{equation}
\label{equa:scibert}
s^{+VeriRel} = f(c,d) \approx s^{combo}_{c,d}
\end{equation}

where $f(c,d)$ is a function that models the relationship between a claim $c$ and a document, using a trainable end-to-end model to learn from calculated $s^{combo}_{c,d}$ and gold evidence.

As shown in Figure~\ref{fig:VeriFED}, the blue pathway generates $s^{combo}_{c,d}$ as training signals from semantic relevance and verification feedback, which are then used to train \textit{+VeriRel}. The orange pathway presents the evaluation of ranked documents (1) by the learned model \textit{+VeriRel}, and (2) by $s^{combo}_{c,d}$ directly (denoted as \textit{ComboScorer}).

\section{Experiments}
This section discusses a series of experiments to (1) address our preliminary study that validates our assumption about the usefulness of verification feedback and (2) train and evaluate our proposed \textit{+VeriRel} reranking model in a scientific fact checking system. Our code and data are publicly available.\footnote{https://github.com/xingyu-deng/VeriRel}

\subsection{Datasets}
Three publicly available datasets were used to provide evidence-annotated corpora: 

\noindent \textbf{SciFact}~\citep{wadden-etal-2020-fact} contains 809/300 claims for training/validation and 5,183 scientific abstracts from S2ORC~\citep{lo-etal-2020-s2orc}. SciFact also includes a test set consisting of 300 claims. Gold evidence for these has not been publicly released but can be inferred from the SciFact-Open dataset \citep{wadden-etal-2022-scifact} to produce a dataset consisting of 279 claims which was used to evaluate the approaches.  

\noindent \textbf{SciFact-Open}~\citep{wadden-etal-2022-scifact} is an extension of SciFact which reuses 279 of its test claims while expanding the corpus to 500K abstracts with new evidence annotations.
In this study, a distilled dataset comprising evidence only from newly incorporated documents is used to assess the generalisability.

\noindent \textbf{Check-COVID}~\citep{wang-etal-2023-check-covid} comprises 1,504 COVID-19 related claims and 347 scientific documents. The full dataset is used for evaluation.

All models were trained only using the SciFact training and validation sets, then evaluated using the SciFact, SciFact-Open and Check-COVID evaluation data.

\subsection{Model Selection}
\textbf{Reranker}:  The BM25\cite{robertson1995okapi} and monoT5-3B\cite{nogueira-etal-2020-document} pipeline was adopted as a baseline since it was the best-performing retrieval system for SciFact \cite{liu2025leveraging,sun-etal-2023-chatgpt}, as reported through the BEIR leaderboard \citep{thakur2021beir}, and effective for scientific fact checking \citep{pradeep-etal-2021-scientific, wuhrl-klinger-2021-claim, wadden-etal-2022-multivers,wadden-etal-2022-scifact, vladika-matthes-2023-scientific}. Both monoT5-3B and its MS MARCO MED variant, denoted monoT5-3B(Med), were applied given the biomedical nature of the datasets.

\noindent\textbf{Claim Verifier}:  MultiVerS \citep{wadden-etal-2022-multivers} was used since it is the current state-of-the-art claim verifier on SciFact and SciFact-Open according to the task leaderboard and published results~\citep{wadden-etal-2022-multivers,wadden-etal-2022-scifact}. MultiVerS uses the Longformer model \citep{beltagy2020longformer} which enables the processing of long documents to cover abstract-level text and avoid information loss. MultiVerS is initialised with the checkpoint \cite{wadden-etal-2022-multivers} trained on three datasets: FEVER \citep{thorne-etal-2018-fever}, PubMedQA \citep{jin-etal-2019-pubmedqa} and Evidence Inference \citep{lehman-etal-2019-inferring, deyoung-etal-2020-evidence}.

\subsection{Negative sampling}\label{ssec:negative_sample}

\noindent All documents in the datasets other than gold evidence are unannotated and treated as being labelled {\tt NOT\;ENOUGH\;INFORMATION}. The number of negative samples plays a key role in balancing generalisation and overfitting \citep{wadden-etal-2022-scifact}. In practice, verification models are commonly trained with a substantial number of negative samples to improve robustness and achieve higher overall verification performance, particularly in terms of F1 score \cite{li2021paragraph, wadden-etal-2022-multivers, zhang-etal-2021-abstract, wadden-etal-2020-fact}. For instance, MultiVerS uses 20 negative samples for each positive one during training \cite{wadden-etal-2022-multivers}.
% per instance during training to strike a better balance between precision and recall \cite{wadden-etal-2022-multivers}.

To systematically investigate how verification supervision affects document reranking, the verifier was trained with different numbers of negative documents (N = ${5,10,20}$), randomly sampled from the top-100 documents ranked by BM25 in the SciFact train set. This verifier, used to provide feedback ($s^{v}_{c,d}$), is referred to as V-MultiVerS, to distinguish it from the off-the-shelf MultiVerS used in later verification evaluations. These signals contribute to supervision $s^{combo}_{c,d}$ for \textit{+VeriRel}, allowing examination of how negative sampling strategies affect the quality and generalisability of retrieval training.

Table~\ref{tab:claim_verifier_performance} shows the effect of different negative sampling strategies (N) on verification performance. It can be observed that using more negative documents (N=20) improves precision (i.e., specificity) and achieves higher verification performance (as F1 score), while fewer (N=5) yield better recall and generalisation.

It is possible that using fewer negative samples (i.e. lower values for N) exposes the verifier to more informative borderline cases, leading to richer supervision signals for document reranking. This may help reduce overfitting to trivial non-evidence examples and improve cross-domain generalisation when training \textit{+VeriRel}.

\begin{table}[t]
\caption{Verification performance on SciFact.}
\vspace{-10pt}
\small
\centering
\resizebox{0.35\textwidth}{!}{
\begin{tabular}{lcccc}
\hline
Model            & \#N & \multicolumn{1}{c}{Precision} & \multicolumn{1}{c}{Recall} & \multicolumn{1}{c}{F1} \\ \hline
\multirow{3}{*}{V-MultiVerS}  & 20 & \textbf{62.16}                & 72.52                      & \textbf{66.94}         \\
 & 10 & 57.71                         & 72.52                      & 64.27                  \\
 & 5 & 41.45                         & \textbf{77.48}             & 54.00                  \\ \hline
\end{tabular}
}
\label{tab:claim_verifier_performance}
\vspace{-10pt}
\end{table}

\subsection{+VeriRel Configurations}\label{ssec:model_config}
Training $s^{combo}$ consists of (1) a semantic relevance score ($s^{r}_{c,d}$), which is calculated using a sigmoid function over the output of monoT5-3B, and (2) a verification feedback score ($s^{v}_{c,d}$) from V-MultiVerS.
Different values for the hyperparameter $\alpha$ in Eq.~\ref{ARF_equation} were experimented with. The best-performing was found to vary across negative sampling strategies but values around 0.5 consistently performed well. To ensure fair comparison the simple and intuitive setting of $\alpha$ = 0.5 was adopted for all \#N settings.

To compute $s^{+VeriRel}$ in Eq.~\ref{equa:scibert}, a pre-trained model SciBERT~\citep{beltagy-etal-2019-scibert}, effective in scientific domains~\citep{wadden-etal-2020-fact,kotonya-toni-2020-explainable-automated,tan-etal-2023-multi2claim}, was used to calculate a joint relevance score in the range of [0,1]. During model training, the supervision label of document $d$ is set to 1 if $d$ is gold evidence or $s^{combo}_{c,d}$ otherwise. Model training data consisted of the top 20 documents ranked by $s^{combo}_{c,d}$ in the SciFact train set combined with any gold evidence not otherwise included. 

\subsection{Evaluation Metrics}\label{ssec:evaluation}
\noindent \textbf{Recall@k} assesses the proportion of relevant evidence included in the top k results: 
\begin{equation}
   Recall@k =   \frac{N(retrieved@k)}{N(gold\_evidence)}
\end{equation}

where $N(retrieved @k)$ and $N(gold\_evidence)$ are, respectively, the number of gold evidence documents in the top $k$ of ranked list and in the entire corpus.

\noindent \textbf{Verification performance} is measured using the standard metrics provided from SCIVER shared task (precision, recall and F1) \citep{wadden-lo-2021-overview}. 
The recreated SciFact is denoted as SciFact(offline) to distinguish it from SciFact(leaderboard). For SciFact(leaderboard), we use the unprocessed dataset and evaluate by submitting to the provided leaderboard.\footnote{\label{fn:leaderboard}\url{https://leaderboard.allenai.org/scifact}}

SciFact-Open and Check-COVID are excluded from verification evaluation due to MultiVerS's limited generalisability, as it performs poorly even when providing gold documents in the oracle setting.

\section{Results}
\label{sec:experiment_result}

\renewcommand{\arraystretch}{1.12}
\begin{table*}[ht]
\begin{minipage}{\textwidth}
\caption{Performance of baselines, ComboScorer and \textit{+VeriRel} in Recall@$k$ with cut-off $k$ ranges from 50 to 1.}
\vspace{-5pt}
\resizebox{\textwidth}{!}{
\begin{tabular}{cccccccccccccccccccccc}
\hline
\multirow{2}{*}{Method}            &     &                & \multicolumn{4}{c}{Recall - SciFact}                              &                &  &                & \multicolumn{4}{c}{Recall - SciFact-Open}                         &                &  &                & \multicolumn{4}{c}{Recall - Check-COVID}                          &                \\ \cline{3-8} \cline{10-15} \cline{17-22} 
                                   &     & R@50           & R@20           & R@10           & R@5            & R@3            & R@1            &  & R@50           & R@20           & R@10           & R@5            & R@3            & R@1            &  & R@50           & R@20           & R@10           & R@5            & R@3            & R@1            \\ \hline
\multicolumn{1}{l}{BM25}           &     & 73.68          & 67.94          & 61.24          & 55.50          & 48.33          & 35.41          &  & 59.76          & 43.82          & 31.87          & 23.51          & 16.33          & 7.57           &  & 87.91          & 81.96          & 75.02          & 67.59          & 61.35          & 46.18          \\
\multicolumn{1}{l}{monoT5-3B}      &     & 90.91          & 87.56          & 85.65          & 78.47          & 70.33          & 55.02          &  & 87.25          & 72.11          & 58.96          & 39.44          & 29.88          & 11.16          &  & 95.84          & 93.16          & 89.49          & 82.06          & 74.93          & 58.28          \\
\multicolumn{1}{l}{monoT5-3B(Med)} &     & 91.39          & 87.56          & 85.17          & 78.95          & 70.81          & 55.50          &  & 85.66          & 71.31          & 57.77          & 41.04          & 28.69          & 10.36          &  & 95.54          & 93.26          & 89.20          & 81.86          & 74.93          & 57.88          \\ \hline
\multicolumn{1}{l}{Ours}                               & \#N &                &                &                &                &                &                &  &                &                &                &                &                &                &  &                &                &                &                &                &                \\ \hline
\multirow{3}{*}{ComboScorer}       & 20  & \textbf{92.82} & \textbf{89.47} & 85.65          & 80.38          & \textbf{77.51} & \textbf{61.72} &  & 87.65          & 72.11          & 62.15          & 43.82          & 30.68          & 11.55          &  & 95.74          & 93.76          & 90.68          & 83.75          & 76.51          & 56.39          \\
                                   & 10  & 92.34          & 89.00          & \textbf{87.08} & \textbf{82.30} & 76.56          & 60.29          &  & 88.84          & 74.10          & \textbf{63.75} & 44.22          & 33.07          & 11.16          &  & \textbf{96.13} & 94.05          & \textbf{91.58}          & \textbf{84.64}          & 76.71          & 59.56          \\
                                   & 5   & 92.34          & 88.52          & 85.65          & 81.34          & 73.21          & 55.98          &  & \textbf{91.63} & \textbf{76.49} & 59.36          & \textbf{47.01} & \textbf{35.86} & \textbf{11.95} &  & \textbf{96.13} & \textbf{94.55} & 91.48 & 84.04 & \textbf{77.80} & \textbf{61.84} \\ \hline
baseline ($\alpha=0$)              & --  & 90.91          & 88.52          & 85.65          & 78.95          & 73.21          & 57.89          &  & 84.46          & 70.92          & 55.38          & 35.46          & 24.70          & 9.96           &  & 96.73          & 91.50          & 83.66          & 71.24          & 60.13          & 43.14          \\ \hline
\multirow{3}{*}{+VeriRel}          & 20  & \textbf{91.87} & 89.47          & \textbf{87.08} & 79.90          & 75.12          & 60.77          &  & 85.26          & 70.52          & 56.97          & 40.24          & 27.89          & 9.96           &  & 96.73          & 92.81          & 90.85          & 78.43          & 69.28          & 48.37          \\
                                   & 10  & \textbf{91.87} & 89.47          & 85.65          & 80.38          & 75.12          & 61.72          &  & 86.45          & 71.31          & 58.17          & 40.64          & 27.09          & 10.76          &  & 96.73          & 94.12          & 91.50          & 81.05          & 71.24          & 53.59          \\
                                   & 5   & \textbf{91.87} & \textbf{90.43} & \textbf{87.08} & \textbf{82.30} & \textbf{75.60} & \textbf{62.20} &  & \textbf{87.65} & \textbf{72.91} & 58.57          & \textbf{41.43} & 28.69          & \textbf{11.55} &  & \textbf{97.39} & \textbf{96.08} & \textbf{92.81} & \textbf{86.93} & \textbf{80.39} & 55.56          \\ \hline
\end{tabular}
}

\label{tab:comboscorer_performance_scifact}
\end{minipage}
\vspace{-5pt}
\end{table*}
\renewcommand{\arraystretch}{1.0}

\begin{table}[h]
\caption{Verification performance by inputting document retrieved by baseline and by proposed \textit{+VeriRel} with $N=5$.}
\vspace{-5pt}
\resizebox{0.43\textwidth}{!}{
\begin{tabular}{llccclc}
\hline
\multicolumn{1}{c}{\multirow{2}{*}{\begin{tabular}[c]{@{}c@{}}Model\\ + MultiVerS\end{tabular}}} &  & \multicolumn{3}{l}{Verification performance}     &                      & \multicolumn{1}{l}{Retrieval performance} \\ \cline{3-5} \cline{7-7} 
\multicolumn{1}{c}{}                                                                             &  & P              & R              & F1             & \multicolumn{1}{c}{} & Recall@k                                  \\ \hline
\textbf{Top 10}                                                                                  &  & \multicolumn{3}{c}{SciFact(offline)}            & \multicolumn{1}{c}{} & Recall@10                                 \\ \hline
monoT5-3B                                                                                  &  & 74.52          & 55.98          & 63.93          &                      & 86.60                                     \\
+VeriRel                                                                                     &  & \textbf{75.88} & \textbf{61.72} & \textbf{68.07} &                      & \textbf{88.04}                            \\ \hline
\textbf{Top 5}                                                                                   &  & \multicolumn{3}{c}{SciFact(offline)}            & \multicolumn{1}{c}{} & Recall@5                                  \\ \hline
monoT5-3B                                                                                  &  & \textbf{73.03} & 62.20          & 67.18          &                      & 80.38                                     \\
+VeriRel                                                                                     &  & 72.48          & \textbf{65.55} & \textbf{68.84} &                      & \textbf{84.21}                            \\ \hline
\textbf{Top 3}                                                                                   &  & \multicolumn{3}{c}{SciFact(offline)}            & \multicolumn{1}{c}{} & Recall@3                                  \\ \hline
monoT5-3B                                                                                  &  & 76.43          & 57.42          & 65.57          &                      & 73.68                                     \\
+VeriRel                                                                                     &  & \textbf{77.78} & \textbf{63.64} & \textbf{70.00} &                      & \textbf{78.95}                            \\ \hline
\textbf{Top 10}                                                                                  &  & \multicolumn{3}{c}{SciFact(leaderboard)}         & \multicolumn{1}{c}{} &                                           \\ \hline
monoT5-3B                                                                                  &  & \textbf{73.83} & \textbf{71.17} & \textbf{72.48} &                      & Not accessible                            \\
+VeriRel                                                                                     &  & \textbf{73.83} & \textbf{71.17} & \textbf{72.48} &                      & Not accessible                            \\ \hline
\textbf{Top 5}                                                                                   &  & \multicolumn{3}{c}{SciFact(leaderboard)}         & \multicolumn{1}{c}{} &                                           \\ \hline
monoT5-3B                                                                                  &  & 74.88 & 69.82          & 72.26          &                      & Not accessible                            \\
+VeriRel                                                                                     &  & \textbf{75.12} & \textbf{70.72} & \textbf{72.85} &                      & Not accessible                            \\ \hline
\textbf{Top 3}                                                                                   &  & \multicolumn{3}{c}{SciFact(leaderboard)}         & \multicolumn{1}{c}{} &                                           \\ \hline
monoT5-3B                                                                                  &  & \textbf{77.04} & 68.08          & 72.25          &                      & Not accessible                            \\
+VeriRel                                                                                     &  & 75.86          & \textbf{69.37} & \textbf{72.47} &                      & Not accessible                            \\ \hline
\end{tabular}
}
\label{tab:verification_improvement_by_scibert}
\vspace{-5pt}
\end{table}

Evaluation compares retrieval effectiveness of documents ranked (1) by $s^{combo}_{c,d}$ directly (denoted as ComboScorer) and (2) by a trained ranking model \textit{+VeriRel}, using Recall@k across SciFact, SciFact-Open, and Check-COVID.

Table~\ref{tab:comboscorer_performance_scifact} shows that both approaches outperform the strong baselines, particularly at lower values of $k$. In addition, an ablation study excluded the verification signal by setting $\alpha=0$, denoted by ``baseline ($\alpha=0$)". These results highlight the benefit of incorporating verification feedback into document scoring, either directly at inference or as supervision for reranker training.

On SciFact, both ComboScorer and \textit{+VeriRel} outperform the state-of-the-art monoT5-3B across all Recall@k values.
Among ComboScorer variants, including more negative samples (N=20) yields the highest early retrieval scores (e.g., Recall@1 = 61.72, Recall@3 = 77.51), suggesting that more negatives help improve document-level precision. Interestingly, when ComboScorer is used to supervise \textit{+VeriRel} training, the optimal negative sampling setting shifts: \textit{+VeriRel} (N=5) achieves the best retrieval performance, surpassing all configurations including its own supervision source, with Recall@1 = 62.20, Recall@3 = 75.60 and Recall@5 = 82.30. This contrast highlights a key insight: although high-N verification signals yield stronger precision in isolation, lower-N settings provide more informative supervision for training a general-purpose reranker.

To evaluate generalisability, all approaches were also evaluated using SciFact-Open and Check-COVID. The \textit{+VeriRel} variant trained with N=5 demonstrates the most consistent performance across both datasets, outperforming other configurations at all Recall@k levels. 
Notably, Check-COVID presents a greater challenge due to its distinct domain and evolving terminology, yet \textit{+VeriRel}(N=5) achieves the highest scores with substantial gains over all baselines, indicating strong robustness to distribution shift.
These results suggest that supervision from fewer hard negative documents may enhance generalisability by exposing the model to more diverse and ambiguous training signals. While high-N verifiers offer stronger in-domain signals, they may lead to overfitting to domain-specific patterns, reducing robustness under distribution shift.
One limitation of the Check-COVID dataset is that each claim is annotated with evidence from a single source document, while relevant content in other documents remains unlabelled. This incomplete annotation penalises correct retrievals, particularly at Top-1, where relevant but unlabelled documents are treated as false positives.
As a result, while \textit{+VeriRel} improves overall retrieval performance, its performance on Check-COVID is still underestimated.

For evaluation of downstream verification, documents reranked by \textit{+VeriRel}(N=5) improve MultiVerS performance over monoT5-3B across top-10, top-5, and top-3 settings, as shown in Table~\ref{tab:verification_improvement_by_scibert}.
For instance, on SciFact (offline), F1 improves from 65.57\% to 70.00\% at top-3 input. On the SCIVER leaderboard\hyperref[fn:leaderboard]{\textsuperscript{2}}, \textit{+VeriRel} matches or slightly exceeds monoT5-3B in verification F1 while requiring fewer documents. These gains support that improved reranking quality translates directly to better verification.
In summary, \textit{+VeriRel} improves both document retrieval and downstream verification, demonstrating the benefit of integrating verification feedback into document relevance estimation.

\section{Conclusion}

This study presents \textit{+VeriRel}, a document reranking approach that enhances scientific fact checking by leveraging feedback from verification stages. By integrating a verification reward model into the ranking process, \textit{+VeriRel} prioritises more relevant and evidential documents and consistently improves retrieval accuracy. Experiments demonstrate its strong generalisability, particularly on large, diverse, and previously unseen corpora, exemplified by fast-evolving domains such as COVID-19. The use of downstream verification feedback as an automated relevance feedback mechanism enables more robust and effective document retrieval, which is particularly crucial in scientific domains that demand precision and high-quality evidence.
Results presented here demonstrate the benefit of decoupling the training of the verification reward model and the final claim verifier. The reward model should be trained with fewer negative samples to support generalisable evidence identification. Once the reranker is optimised using this feedback, a separate, task-specific verifier can be trained for inference.
The improvements from integrating verification supervision may transfer to general fact checking. Future work could adapt our training paradigm to open-domain datasets, especially where dependence on search APIs limits retrieval quality and transparency.

\section*{GenAI Usage Disclosure}
This manuscript has benefited from the use of Generative AI (ChatGPT) to improve the quality of text produced by the authors. The authors remain responsible for the content.

\bibliographystyle{ACM-Reference-Format}
\bibliography{sample-base}

\end{document}